\documentclass[preprint]{aastex63}
\usepackage{amssymb}
\usepackage{amsmath}
\usepackage{ulem}
\usepackage{graphicx}

\newcommand{\FeHeq}[1]{$\mbox{[Fe/H]}={#1}$}    

\received{}
\revised{}
\accepted{}
\submitjournal{ApJ}

\shorttitle{r-process-enhanced stars}
\shortauthors{Xing et al.}

\begin{document}

\title{Detection of the actinide Th in an r-process-enhanced star with accretion origin
}

\correspondingauthor{Qianfan Xing, Gang Zhao}
\email{qfxing@nao.cas.cn, gzhao@nao.cas.cn}

\author[0000-0003-0663-3100]{Qianfan Xing}
\affiliation{Key Lab of Optical Astronomy, National Astronomical
  Observatories, Chinese Academy of Sciences (CAS) \\
A20 Datun Road, Chaoyang, Beijing 100101, China}

\author[0000-0002-8980-945X]{Gang Zhao}
\affiliation{Key Lab of Optical Astronomy, National Astronomical
  Observatories, Chinese Academy of Sciences (CAS)  \\
A20 Datun Road, Chaoyang, Beijing 100101, China}
\affiliation{Schoold of Astronomy and Space Science, University of Chinese Academy of
Sciences \\
No.19(A) Yuquan Road, Shijingshan District, Beijing, 100049, China}

\author[0000-0002-8975-6829]{Wako Aoki}
\affiliation{National Astronomical Observatory of Japan \\ 2-21-1 Osawa,
  Mitaka, Tokyo 181-8588, Japan
}
\affiliation{Astronomical Science Program, Graduate Institute for Advanced Studies, SOKENDAI, 2-21-1 Osawa, Mitaka, Tokyo 181-8588, Japan}

\author[0000-0002-0389-9264]{Haining Li}
\affiliation{Key Lab of Optical Astronomy, National Astronomical
  Observatories, Chinese Academy of Sciences (CAS)  \\
A20 Datun Road, Chaoyang, Beijing 100101, China}

\author[0000-0003-2868-8276]{Jingkun Zhao}
\affiliation{Key Lab of Optical Astronomy, National Astronomical
  Observatories, Chinese Academy of Sciences (CAS)  \\
A20 Datun Road, Chaoyang, Beijing 100101, China}

\author[0000-0002-8077-4617]{Tadafumi Matsuno}
\affiliation{National Astronomical Observatory of Japan \\ 2-21-1 Osawa,
  Mitaka, Tokyo 181-8588, Japan
}
\affiliation{Astronomical Science Program, Graduate Institute for Advanced Studies, SOKENDAI, 2-21-1 Osawa, Mitaka, Tokyo 181-8588, Japan}
\affiliation{Astronomisches Rechen-Institut, Zentrum für Astronomie der Universität Heidelberg, Mönchhofstraße 12-14, 69120 Heidelberg, Germany}

\author[0000-0002-4318-8715]{Takuma Suda}
\affiliation{Department of Liberal Arts, Tokyo University of Technology, Nishi Kamata 5-23-22, Ota-ku, Tokyo 144-8535, Japan}
\affiliation{Research Center for the Early Universe, The University of Tokyo, 7-3-1 Hongo, Bunkyo-ku, Tokyo 113-0033, Japan}

\begin{abstract}
The thorium and six second-peak r-process element ($56\leq$ Z $\leq72$) abundances are determined for the $\alpha$-poor star LAMOST J1124+4535 based on a high-resolution spectrum obtained with the High Dispersion Spectrograph (HDS) on the Subaru telescope. The age of J1124+4535 is 11.3 $\pm$ 4.4 Gyr using thorium and other r-process element abundances. J1124+4535 is confirmed to be a Galactic halo metal-poor (\FeHeq{-1.27}$\pm$0.1) star with extreme r-process element over-abundance ([Eu/Fe] = 1.13$\pm$0.08) and $\alpha$ element deficiency ([Mg/Fe] = -0.31$\pm$0.09) by the LAMOST-Subaru project. Along with the sub-solar $\alpha$ to iron ratios (e.g. [Mg/Fe], [Si/Fe], [Ca/Fe]), the relatively low abundances of Na, Cr, Ni and Zn in J1124+4535 show significant departure from the general trends of the Galactic halo but are in good agreement with those of dwarf galaxies. The chemical abundances and kinematics of J1124+4535 suggest it was formed in the late stage of star formation in a dwarf galaxy which has been disrupted by the Milky Way (MW). The star formation of its progenitor dwarf galaxy lasted more than 2 Gyr and has been affected by a rare r-process event before the occurrence of accretion event.

\end{abstract}




\section{Introduction}\label{sec:intro}

The rapid neutron-capture (r-) and slow neutron-capture (s-) processes are mainly responsible for the nucleosynthesis of the elements heavier than iron. For decades, the origin of r-process elements and the main astrophysical site of r-process are still debated \citep{Shen2015ApJ,van2015MNRAS,van2020MNRAS,Cescutti2015AA,Safarzadeh2019ApJ,Haynes2019MNRAS}. The proposed viable site for production of r-process elements include neutron star mergers (NSMs) and subsets of core-collapse supernovae (CCSNe), including CCSNe driven by neutrino-driven wind \citep{Wanajo2018ApJ} and magneto-rotational instability SNe \citep{Winteler2012ApJ}. Most notably, the gravitational wave event GW170817 \citep{Abbott2017PhRvL} is confirmed to be~merging of two neutron stars in a binary system. The follow-up observations to obtain light curve and spectra for this event have identified~r-process elements (Sr and lanthanides) in the ejecta of NSM \citep{Watson2019Natur,Chornock2017ApJ,Domoto2022ApJ}, providing strong supports for NSM as the most promising r-process site.~However, it is not clear whether NSMs could occur frequently enough at the early stage of the MW and produce sufficient r-process yields for the  r-process enrichment observed in very metal-poor stars without other site of r-process \citep{Argast2004AA}.

The records of the nucleosynthesis of heavy elements via r-process events are imprinted on the chemical abundances of long-lived metal-poor stars. A great deal of observational research (e.g. \citealt{Hansen2018ApJ,Honda2004ApJ,Zhao1991AA}) has been performed for understanding the main astrophysical site of r-process by analyzing the detailed chemical abundances of metal-poor stars with extreme r-process enhancement, such as so-called r-II stars \citep{Beers2005ARAA} that have Eu/Fe ratios higher than solar ratios by more than 10 times ([Eu/Fe] $> +1.0$)\footnote{
The standard notations
[X/Y]$=\log(N_{\rm X}/N_{\rm Y})-\log(N_{\rm X}/N_{\rm Y})_{\odot}$ and $\log A({\rm X})=\log(N_{\rm X}/N_{\rm H})+12$ for elements X and Y are adopted in this work.}. With the effort of the R-Process Allicance (RPA; \citealt{Hansen2018ApJ}) project, the number of r-II star has increased to over 60. The studies of chemical abundances of such stars like CS 22892-052 (\citealt{Sneden1994ApJ}; [Eu/Fe] = +1.6 and [Fe/H] = -3.1) have revealed that the abundances pattern of the heavy neutron-capture elements (at least for Ba to Hf) in r-II stars from the Galactic halo and satellite dwarf galaxies \citep{Aoki2007PASJ,Ji2016ApJ} agree well with that of the solar system r-process component, suggesting the r-process pattern is universal and may result from a dominating astrophysical site.

Several theoretical research aimed to test whether NSM could be the main site of r-process by comparing the Eu abundance trend in Galactic halo stars with simulation results from galactic chemical evolution (GCE) models. Early studies (e.g. \citealt{Argast2004AA}) found that the NSM could not well explain the distribution and scatter of [Eu/Fe] vs. [Fe/H], because the timescale of mergers of binary neutron stars is longer than that of metal enrichment in the early Galaxies in these models. However, recent studies have revealed that the satellite dwarf galaxies around MW are important building blocks for the formation of the Galactic halo \citep{Helmi2020ARAA,Zhao2021SCPMA,Zhao2009ApJ}. The contribution of the dwarf galaxies should be considered in the theoretical analysis for the Eu abundance trend in the halo \citep{Tsujimoto2014AA,Ishimaru2015ApJ}. In recent years, dozens of r-I and r-II stars have been found in stellar streams (e.g. \citealt{Hansen2021ApJ}), classical dwarf galaxies, and even in ultra faint dwarf galaxies (UFD). \citet{Ji2016ApJ} have discovered 7 r-rich stars with [Fe/H] $< -2$ in Reticulum II (Ret II) and concluded a rare r-process event such as NSM is responsible for the r-process enhancement in this UFD. Since most of r-II stars found in the early researches have metallicities in a relatively narrow region ([Fe/H] $< -2.5$ ; \citealt{Sneden2008ARAA}), overlapping with the metallicities of UFD stars, UFDs, such as Ret II, are suggested to be the source of r-rich stars in the Galactic halo. This scenario is supported by the kinematic analysis of 35 r-rich halo stars by \citet{Roederer2018AJ}, which showed that most of them do not follow the Galactic rotation and are clustered in the phase space and metallicity. Furthermore, \citet{Roederer2018ApJ} and \citet{Xing2019NatAs} have found two r-II stars in medium metal-poor ([Fe/H] $\sim -1.4$) halo stars. Their metallicities are higher than the r-rich stars in UFDs, but in consistent with r-rich stars in classical dwarf galaxies, such as Ursa Minor (UMi) and Fornax \citep{Reichert2021ApJ}. Notably, the r-II star found by \citet{Xing2019NatAs} exhibit sub-solar [X/Fe] ratios for $\alpha$ elements (e.g. Mg, Si, Ca, and Ti) at [Fe/H] $< -1$, which are well-known chemical labels for identifying accreted halo stars (e.g. \citealt{Helmi2020ARAA}). Similar $\alpha$-poor stars with extreme r-process enhancements have been identified by the following high-resolution spectroscopic observations of the stars in the Galactic halo (e.g. \citealt{Ezzeddine2020ApJ,Sakari2019ApJ}) and satellite dwarf galaxies (e.g. \citealt{Reichert2021ApJ}), suggesting dwarf galaxies are potential source of r-rich halo stars.

In this paper we obtain the abundances of 12 neutron-capture elements for exploring the abundance pattern of neutron-capture elements for J1124+4535 \citep{Xing2019NatAs}, along with measurement of age from Th abundance. J1124+4535 is selected from the LAMOST survey as candidate $\alpha$-poor stars. We have performed twice follow-up high-resolution spectroscopic observations with different resolution and wavelength ranges for J1124+4535 as part of observation task in the LAMOST-Subaru project \citep{Aoki2022ApJ,Li2022ApJ}. Based on the first spectrum (R $\sim 45,000$) obtained with Subaru/HDS on 2017 February 27, our previous work \citep{Xing2019NatAs} confirmed J1124+4535 is a Galactic halo star with extremely enhanced r-process element abundances and sub-solar [$\alpha$/Fe] ratio. The spectrum with higher resolution (R $\sim 60,000$) used in this paper was obtained on 2019 June 9, covering 3500 to 5250 \AA~wavelength range. The determination of chemical abundances using those spectra is presented in Section 2. In Section 3 we discuss the r-process pattern and measurement of age from the Th abundance. The kinematic analysis based on the updated parallax and proper motions from Gaia DR3 \citep{Gaia2023AA} is also included in Section 3.

\section{Observations and data analysis} \label{sec:data}

LAMOST J1124+4535 was identified as a metal-poor star to have a relatively low Mg abundance from a sample of candidate $\alpha$-poor stars \citep{Xing2015RAA} observed by LAMOST spectroscopic survey \citep{Zhao2006ChJAA}. All of these candidates were found to have sub-solar [Mg/Fe] based on spectrum synthesis analysis of Mg Ib lines from spectra with a resolving power $R \sim 1800$ \citep{Xing2014ApJ}.

A high-resolution spectroscopic observation of J1124+4535 took place on 2017 February 27 using HDS on the Subaru Telescope \citep{Noguchi2002PASJ}. The high-resolution spectrum covers the wavelength range 4000-6800 \AA~with a gap of 5330-5430 \AA. The resolving power R $\sim 45,000$ is obtained using a 0$^{''}$.8 slit and 2 x 2 CCD pixel binning. The signal-to-noise ratios (S/N) per pixel at 4300 \AA~and 5000 \AA~are 50 and 70, respectively. Based on this spectrum, \citet{Xing2019NatAs} revealed the r-process enhancement and deficiencies of Na and $\alpha$ elements in J1124+4535 with its metallicity of [Fe/H] $= -1.27$. The highly enhanced Eu abundance ([Eu/Fe] $> 1$) and sub-solar [Mg/Fe] make this star to be the first known $\alpha$-poor r-II star.

In order to obtain detailed r-process pattern of J1124+4535, we observed J1124+4535 again with Subaru/HDS on 2019 June 9 with 90 minutes exposure. The high-resolution spectrum covers the wavelength range 3500-5250 \AA~with a resolving power of $R \sim 60,000$ using a 0$^{''}$.6 slit and 2 x 2 CCD pixel binning. The S/N ratios in the continuum range are 80 per pixel at 4000 \AA~and 100 per pixel at 4500 \AA. The portion of the spectrum of J1124+4535 covering the absorption line of Th at 4019 \AA~is presented in the top panel of Figure 1. The standard data reduction was performed with Subaru/HDS pipeline\footnote{https://www.naoj.org/Observing/Instruments/HDS/} based on IRAF\footnote{IRAF is distributed by the National Optical Astronomy Observatories, which is operated by the Association of Universities for Research in Astronomy, Inc. under cooperative agreement with the National Science Foundation.}, including overscan subtraction, bias subtraction, linearity correction, cosmic-ray rejection, scattered light subtraction, spectrum extraction and wavelength calibration. The radial velocity (RV) of J1124+4535 is measured from Fe I lines that are used for abundance analysis with equivalent width (EW) measurements. The heliocentric RV derived from the high-resolution spectra, 54.8$\pm$0.5 km s$^{-1}$, is in good agreement with that derived from the LAMOST low-resolution spectrum (57.1$\pm$2.5 km s$^{-1}$), indicating there is no evidence of RV variations caused by unknown companion. The EWs are measured with Splot task in IRAF package by fitting Gaussian or Voigt line profiles relative to the normalized continuum. We compare the EWs of more than 50 lines in common between the two high-resolution spectroscopic observations and find no obvious difference. The mean EW difference is 4.5 m\AA. We here adopt the stellar parameters (T$_{\rm eff}$ = 5180 K, log $g$ = 2.7, $v_{\rm t}$ = 1.5 km s$^{-1}$) and iron abundance ([Fe/H] = -1.27) determined from the previous high-resolution spectroscopic observation as reported in \citet{Xing2019NatAs}. The basic information of J1124+4535 are given in Table 1.

\begin{deluxetable}{llrc}
\tablecolumns{4}
\tablewidth{3pt}
\tabletypesize{\scriptsize}
\tablecaption{Stellar Parameters and kinematics}
\tablehead{\colhead{Parameter} &
 \colhead{Symbol} &
 \colhead{Value} &
 \colhead{References}  \\
 }
\startdata
RA   &    		 $\alpha$ (J2000)  &  11:24:56.61                                     & Gaia DR3 \\
DEC &    		 $\delta$ (J2000)  &   45:35:31.32                                       &  Gaia DR3 \\
Parallax    &       $\varpi$              &   0.09 $\pm$ 0.02 mas                             &  Gaia DR3  \\
Distance  &       $D$                     &   7653.22852$^{+1083.5}_{-838.5}$ pc   & Bailer-Jones et al. 2021 \\
Proper motion ($\alpha$)    & PMRA  &  -0.83 $\pm$ 0.01  mas yr$^{-1}$      & Gaia DR3 \\
Proper motion ($\delta$)     & PMDEC  & -6.72 $\pm$ 0.01 mas yr$^{-1}$     & Gaia DR3 \\
\hline
Effective temperature       & T$_{\rm eff}$  &  5180 $\pm$ 100 K & Xing et al. 2019 \\
Surface gravity       &   log $g$      &   2.7 $\pm$ 0.3 cgs   & Xing et al. 2019 \\
Microturbulent velocity      &   $v_{\rm t}$ & 1.5 $\pm$ 0.3 km s$^{-1}$   &  Xing et al. 2019 \\
Metallicity                          &   [Fe/H]        &  -1.27 $\pm$ 0.1  &  Xing et al. 2019 \\
\hline
Orbital pericentric radius   &   R$_{\rm peri}$   &  4.2$^{+1.42}_{-0.48}$ kpc   &  this study \\
Orbital apocentric radius   &   R$_{\rm apo}$   &  13.5$^{+0.94}_{-0.73}$ kpc &  this study \\
Maximum height   &   Z$_{\rm max}$   &   13.1$^{+0.17}_{-1.29}$  kpc &  this study \\
Eccentricity    &     $e$   &  0.53$^{+0.02}_{-0.09}$  &  this study
\enddata
\end{deluxetable}

Based on the newly obtained high-resolution spectrum covering the wavelength range from 3500 \AA~to 5250 \AA, we are able to determine abundances for 12 new elements heavier than Zn, including Mo I, Ru I, Rh I, Ho II, Er II, Tm II, Yb II, Lu II, Hf II, Os I, Ir I and Th II. The Mo I line at 3864 \AA~was used to determine the Mo abundance, and the Ru I line at 3798 \AA~was identified for measuring Ru abundance. The Rh abundance was derived from the 3700 and 3958 \AA~lines, while Ho abundance was derived from 3810, 3890, and 4045 \AA~lines. The three Er II lines at 3692, 3729, and 3830 \AA~were used to estimate the abundance of Er. The Tm, Yb, and Lu abundances were determined using the 3701 and 3795 \AA~Tm II lines, the 3694 \AA~Yb II line, and the 3507 \AA~Lu II line, respectively. The 3719 and 4093 \AA~Hf II lines, the 4135 \AA~Os I line, and the 3800 \AA~Ir I line were also used to determine the abundances. Chemical abundances for these elements were determined from EW measurements and spectral synthesis using the 2017 version of MOOG \citep{Sneden1973PhDT} and the 1D plane-parallel model atmospheres \citep{Castelli&Kurucz2003IAUS} under the assumption of local thermodynamic equilibrium (LTE). The excitation potential and oscillator strength for lines used for abundance determination are adopted from \citet{Ivans2006ApJ} and \citet{Roederer2018ApJ}, while atomic data and isotopic ratios from \citet{Sneden2008ARAA} and \citet{Kurucz1995} are employed for spectral synthesis of lines with hyperfine structure and isotopic splitting.

Th abundance is derived from a single line at 4019 \AA. This line is blended by CH, Fe, Ni and Ce. We make use of the spectrum synthesis method and the line list in \citet{Mashonkina2014AA} to determine the Th abundance, considering the affection introduced by these blends. As shown in the bottom panel of Figure 1, the abundance log $\epsilon$(Th)= $-0.19$ is adopted by the best fitting synthetic spectrum. The synthetic spectra with abundance variations of $\pm 0.2$ are also shown for comparison. Another Th II line at 3741 \AA~is also detected in the spectrum. The Th abundance derived from this line is 0.05 dex higher than that from Th II 4019 \AA. Taking account of the weakness of the lines and severe blending with Ti I and Sm II lines, we adopt log $\epsilon$(Th)= $-0.19\pm0.09$ from the Th II line at 4019 \AA~for age determination. The error in the Th abundance is composed of fitting error and systematic uncertainties due to changes in stellar parameters. The feature of U line at 3859 \AA~is used to derive the upper limit for U (log $\epsilon$(U) $< -0.85$) based on the spectrum synthesis method. The abundance ratios for J1124+4535 derived from the Subaru/HDS spectrum are listed in Table 2. The abundance of Fe I and solar abundances from literature \citep{Asplund2009ARAA} are adopted for obtaining [X/Fe]. The uncertainties in measurements and stellar parameters are considered for the abundance uncertainties. The total error ($\sigma_{total}$) in Table 2 is the quadratic sum of the random error due to the uncertainty of measurement and systematic errors ($\sigma_{sys}$) due to the stellar parameter uncertainties ($\delta$$T_{\rm eff}$ = 100 K, $\delta$log g = 0.3 dex, and $\delta$$v_{\rm t}$ = 0.3 km s$^{-1}$). The random error is estimated to be $\sigma$$_{\upsilon}$$N$$_{\rm lines}^{-1/2}$, where $\sigma$$_{\upsilon}$ is the dispersion around the mean abundance obtained from different lines for each species, and $N$$_{\rm lines}$ is the number of lines used to derive abundance. The $\sigma$$_{\upsilon}$ of Fe I is adopted for species with $N$$_{\rm lines}$ $<$ 3. For elemental abundances measured by the spectrum synthesis method, the errors are estimated based on comparison with synthetic spectra ($\Delta$$\chi^2$ $= 1$).

\begin{figure*}[h!]
\epsscale{1.1}
\plotone{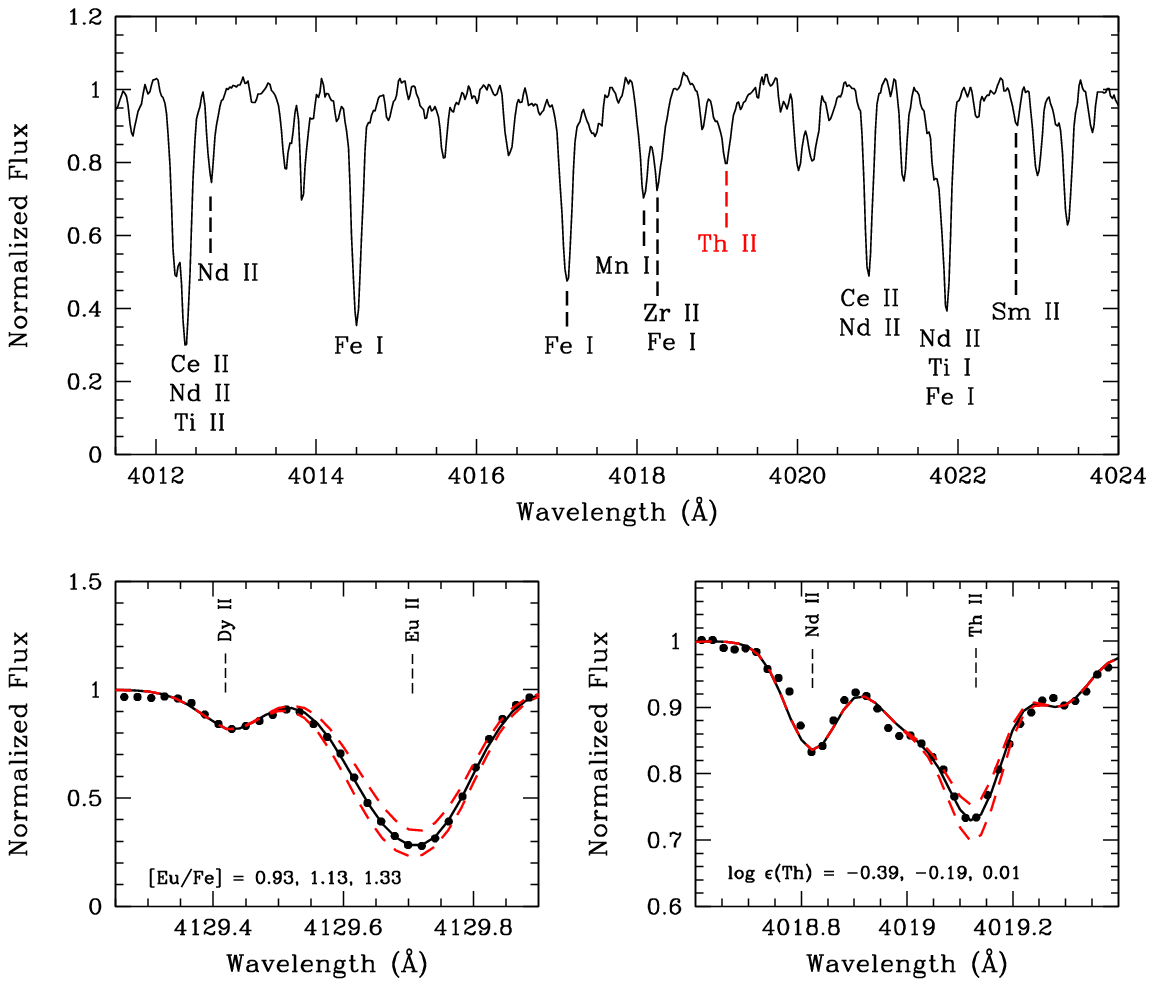}
\caption{Top panel: Selected region of the spectrum of J1124+4535 covering the Th line at 4019 \AA. Bottom panel: The Eu and Th abundances are determined by spectrum synthesis method. The isotope ratios from the r-process component of the Solar System \citep{Arlandini1999ApJ} are adopted for Eu. The observed spectra are plotted as filled circles. The synthetic spectra with best-fit abundances are shown in solid lines. The red dashed lines are synthetic spectra with abundance variations of $\pm0.2$ dex.}
\end{figure*}

\begin{table}
\centering
\caption{Chemical abundances of J1124+4535}
\label{table:abunds}
\begin{tabular}{lcccrcc}
\hline
Element X & $N_{\rm lines}$ & log $\epsilon$ & $\sigma_{sys}$ & $\sigma_{total}^{a} $ & [X/Fe]  & $\sigma_{total}$ \\
\hline
Co \scriptsize{I} & 6 & 3.70 & 0.20 & 0.21 &$-$0.03  & 0.08 \\
Mo \scriptsize{I} & 1 & 1.01 & 0.16 & 0.16 & 0.15  & 0.12 \\
Ru \scriptsize{I} & 1 & 1.25 & 0.16 & 0.17 & 0.77  & 0.14 \\
Rh \scriptsize{I} & 2 & 0.57 & 0.15 & 0.15 &  0.93   & 0.13 \\
Ba \scriptsize{II} & 5 & 1.24 & 0.15 & 0.16 & 0.24  & 0.12 \\
Eu \scriptsize{II} & 1 & 0.38 & 0.11 & 0.12 & 1.13  & 0.08 \\
Ho \scriptsize{II} & 3 & 0.30  & 0.17 & 0.18 & 1.09  & 0.16 \\
Er \scriptsize{II} & 3 & 0.58  & 0.15 & 0.16 &   0.93    & 0.12 \\
Tm \scriptsize{II} & 2 & -0.24 & 0.13 & 0.14 & 0.93     & 0.12 \\
Yb \scriptsize{II} & 1 & 0.37   & 0.19 & 0.20 &  0.80   & 0.16 \\
Lu \scriptsize{II} & 1 & -0.16  & 0.15 & 0.16 &   1.01   & 0.12 \\
Hf \scriptsize{II} & 2 & 0.29   & 0.12 & 0.13 &  0.71    & 0.09 \\
Os \scriptsize{I}  & 1 & 1.19   & 0.14 & 0.14 &  1.06    & 0.10 \\
Ir \scriptsize{I}  & 1 & 1.11   & 0.13 & 0.14 &  1.0    & 0.10 \\
Pb \scriptsize{I} & 1 & $<$ 1.4 & $\cdots$ & $\cdots$ & $<$ 0.92 & $\cdots$ \\
Th \scriptsize{II} & 1 & -0.19 & 0.14 & 0.15 & 1.06  & 0.09 \\
U \scriptsize{II} & 1 & $<$ -0.85  & $\cdots$ & $\cdots$ &  $<$ 0.96 & $\cdots$ \\
\hline
\end{tabular}
\tablenotetext{a}{The total error is the quadratic sum of random error and systematic error ($\sigma_{sys}$) due to the uncertainties on the stellar
parameters ($T_{\rm eff}$, log g, and $v_{\rm t}$). Errors in log $\epsilon$ and [X/Fe] are presented separately.}
\end{table}

\section{Discussion}
\subsection{R-process pattern}

Figure 1 illustrates the highly enhanced Eu ([Eu/Fe] = 1.13$\pm0.1$) in J1124+4535 with the best fitting synthetic spectrum. Such high enhancements of Eu with respect to Fe are usually found in very metal-poor stars ([Fe/H] $< -2$), and are very rare in metal-poor stars with [Fe/H] $> -1.5$, making J1124+4535 to be one of the two known moderately metal-poor r-II stars in the MW field halo stars. The other one is HD 222925 ([Eu/Fe] = +1.33, [Fe/H] = -1.47) discovered by \citet{Roederer2018ApJ}. The abundance pattern of neutron-capture elements measured for J1124+4535 is shown in Figure 2 compared with the scaled solar r-process pattern. The r-process pattern in solar system is determined by the abundance residual of solar abundances from which the s-process abundances are subtracted \citep{Arlandini1999ApJ}.

The abundances in J1124+4535 agree well with the scaled solar r-process pattern (see Figure 2), in consistent with that of other r-II stars. The mean residual is 0.07 dex with a standard deviation of 0.08 dex, smaller than the measured abundance uncertainties for r-process elements. This agreement is prominent for the second-peak elements (56 $\leq$ Z $\leq$ 72) and extends to the third-peak element (76 $\leq$ Z $\leq$ 88), as well as the actinide element Th. However, the abundances of light neutron-capture elements Sr, Y, Zr, Mo, Ru and Rh (first r-process peak elements) in J1124+4535 are significantly lower than that of solar r-process pattern, in contrast to the heavier neutron-capture elements. Similar deficiencies of the first-peak elements have been widely found in other r-process-enhanced stars from the Galactic halo and dwarf galaxies (see the bottom panel of Figure 2). A possible explanation is that the light neutron-capture elements, such as Sr, Y and Zr in the solar system material, are mainly produced by the weak r-process \citep{Siqueira2014AA} rather than the main r-process. The so-called weak r-process cannot synthesize elements heavier than the first-peak elements (38 $\leq$ Z $\leq$ 48).

The actinide Th abundance in J1124+4535 ([Fe/H] = -1.27) is in good agreement with the solar r-process pattern (see Figure 2), similar to the other r-II star HD 222925 with [Fe/H] = -1.47. The log $\epsilon$(Th/Eu) abundances (around -0.5) of these two stars are common among the majority of r-II stars. A noteworthy dispersion among log $\epsilon$(Th/Eu) has been found in very metal-poor and extremely metal-poor stars. Such dispersion is mainly caused by ``actinide boost'' stars with highly enhanced Th \citep{Hill2002AA}. The origin of these stars are still unknown and more analysis of Th abundances for metal-poor stars are required.

\begin{figure*}[h!]
\epsscale{1.1}
\plotone{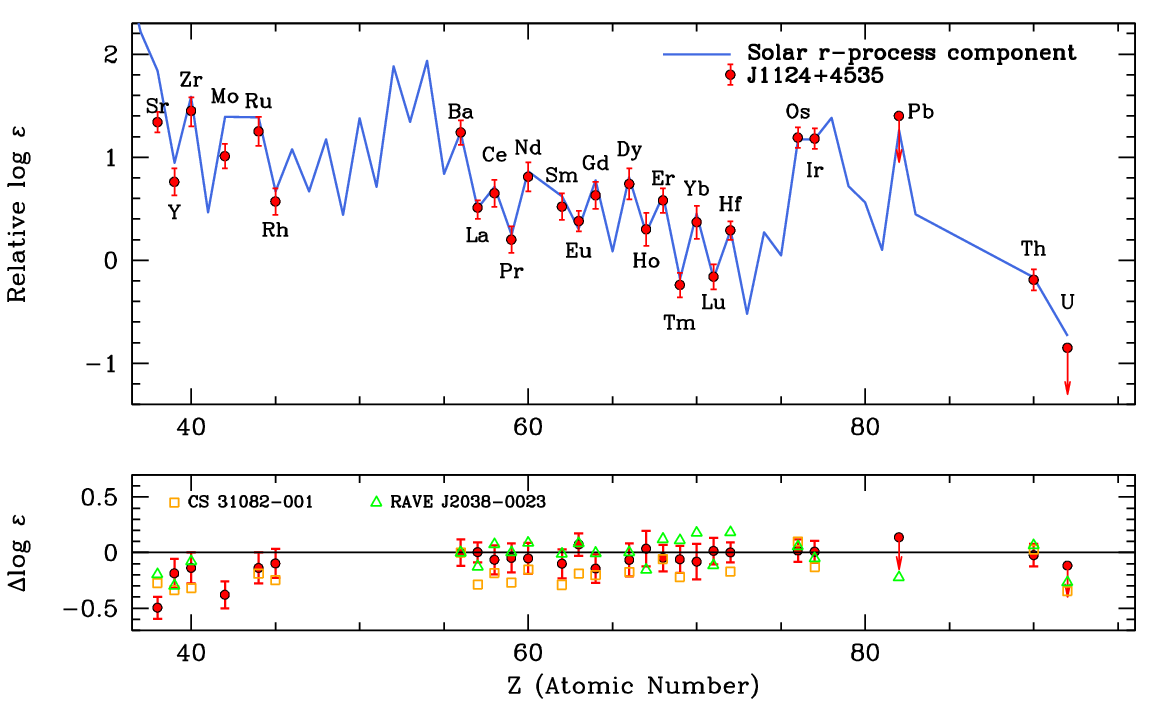}
\caption{Top panel: neutron-capture elemental-abundance pattern for J1124+4535 in comparison with a scaled solar r-process pattern \citep{Arlandini1999ApJ}, normalized to Ba. Bottom panel: differences between the observed abundances and the scaled solar r-process pattern for J1124+4535, CS 31082-001 \citep{Hill2002AA,Sneden2008ARAA} and RAVE J2038-0023 \citep{Placco2017ApJ}.}
\end{figure*}

\subsection{Determination of Age}

Th has a half-life of 14.05 Gyr and allows us to determine the age of J1124+4535 through radioactive-decay dating, with assumption that the r-process material is produced by a single r-process event soon before the formation of J1124+4535. Based on the log $\epsilon$(Th/r) derived from the observed spectrum and initial produce ratios (such as log $\epsilon$(Th/Eu)$_{initial}$ = -0.33 $\pm$ 0.12) given by \citet{Schatz2002ApJ}, the age of the r-process event and J1124+4535 is calculated by

\begin{equation*}\text{t} = 46.67 [\log \epsilon(\rm{Th/r})_{\rm initial} - \log \epsilon(Th/r)_{\rm now}]~Gyr    ~~~~~~ (1)
\end{equation*}

where $\log \epsilon(\rm{Th/r})_{\rm initial}$ is an initial production ratio, while $\log \epsilon(\rm {Th/r})_{\rm now}$ is the observed value after the decay of radioactive Th produced by the r-process event. The resulting age is very sensitive to the uncertainties of the elemental abundances. So we only use stable r-process elements with relatively low uncertainties for the measurement of age. The observed Eu, Hf, Os and Ir abundances are derived by the spectrum synthesis method and have relatively small errors. We make use of abundance ratios Th/Eu, Th/Hf, Th/Os and Th/Ir for age estimates, yielding 11.2 $\pm$ 4 Gyr, 13.1 $\pm$ 4.2 Gyr, 10.3 $\pm$ 4.4 Gyr and 10.7 $\pm$ 4.4 Gyr, respectively. The age uncertainties resulting from abundance uncertainties are listed in Table 3. The average age of 11.3 Gyr is very close to the age determined from Th/Eu. The uncertainty of the adopted age is 4.4 Gyr including the abundance uncertainty and the dispersion around the mean age. Finally, we adopt an age of 11.3 $\pm$ 4.4 Gyr for J1124+4535 by averaging the ages determined from Th/Eu, Th/Hf, Th/Os and Th/Ir ratios. The uncertainties of the production ratios, such as 0.12 dex for Th/Eu chronometer in \citet{Schatz2002ApJ}, are not included on the uncertainties of the age. They could result in an age uncertainty of 5.6 Gyr.

The upper limit of U abundance of J1124+4535 can be used to determine a lower limit on the age of this star by

\begin{equation*}\text{t} = 21.8 [\log \epsilon(\rm{U/Th})_{\rm initial} - \log \epsilon(U/Th)_{\rm now}]~Gyr      ~~~~~~ (2)
\end{equation*}

where $\log \epsilon(\rm{U/Th})_{\rm initial}$ is an initial production ratio, while $\log \epsilon(\rm {U/Th})_{\rm now}$ is the observed value. A lower limit on the age of 9.6 Gyr has been derived using the initial production ratio ($log \epsilon(U/Th)_{initial}$ = -0.22 $\pm$ 0.10) from \citet{Schatz2002ApJ}. The derived lower limit age is consistent with the age determined from the observed Th/r ratios.

The relatively low abundances of Na, Cr, Ni, Zn and $\alpha$ elements in J1124+4535 exhibit significant departure from the general trends of the Galactic halo but are in good agreement with abundance trends of UMi dwarf galaxy (see Figure 3 in \citealt{Xing2019NatAs} for detail), suggesting J1124+4535 was accreted from a disrupted dwarf galaxy which is similar to the surviving dwarf galaxy UMi. The r-II star UMi COS 82 \citep{Aoki2007PASJ} has been found in UMi dwarf galaxy. Its metallicity ([Fe/H] = -1.42) and Eu abundance ([Eu/H] = -0.18) both are close to those of J1124+4535 ([Fe/H] = -1.27 and [Eu/H] = -0.14). On the other hand, we could calculate the mass of progenitor stellar system for J1124+4535 based on its Eu abundance and r-process mass yielded by a r-process event \citep{Tsujimoto2017ApJ}, assuming the r-process material in J1124+4535 are produced by a single event. It turns out the Eu abundance of J1124+4535 can be explained by ejecting of 10$^{-4.5 \pm 1}$M$_\odot$ Eu into a progenitor system with a stellar mass $\sim$ 10$^5$M$_\odot$. The progenitor system could be an independent dwarf galaxy or part of a large one, reaching up to a mass of $\sim$ 10$^6$M$_\odot$. The resulting mass of the progenitor system is similar to the mass of UMi dwarf galaxy (2.9x10$^{5}$M$_\odot$, \citealt{McConnachie2012AJ}). In addition, \citet{Cohen2010ApJ} concluded the star formation of UMi dwarf galaxy lasted about 3 Gyr. The stars formed at the late phase (about 11 Gyr ago) in UMi dwarf galaxy exhibit sub-solar [$\alpha$/Fe] ratio due to the contribution of SN Ia (\citealt{Helmi2020ARAA,Xing2015ApJ}), as shown in Figure 3. This closely matches with the case of J1124+4535, which has a sub-solar [$\alpha$/Fe] ratio and age of 11.3 $\pm$ 4.4 Gyr.

\begin{table}
\centering
\caption{Abundance and Age Uncertainties}
\label{table:Abundance Uncertainties}
\begin{tabular}{lcccccc}
\hline
  & $\Delta$T$_{\rm eff}$ & $\Delta$log $g$ & $\Delta$$v_{\rm t}$ & $\sigma_{random}$ & $\sigma_{total}$ & $\sigma_{age}$ \\
\hline
Th/Eu & 0.01 & 0.03 & 0.04 & 0.07 & 0.0866 & 4.0 \\
Th/Hf & 0.01 & 0.01 & 0.04 & 0.08 & 0.0906 & 4.2 \\
Th/Os & 0.04 & 0.06 & 0.01 & 0.06 & 0.0943 & 4.4 \\
Th/Ir & 0.05 & 0.05 & 0.01 & 0.06 & 0.0933 & 4.4 \\
\hline
\end{tabular}
\end{table}

\subsection{Kinematics}

As the parallax of J1124+4535 provided by Gaia DR2 has an error over 40$\%$, we could not derive any conclusion from the kinematic analysis in our previous work \citep{Xing2019NatAs}. The Gaia DR3 has been released on 2022 June 13 and updated the parallax for J1124+4535. The new parallax ($\varpi$ = 0.09 $\pm$ 0.02 mas) has an error smaller than 25$\%$, resulting a more accurate distance of 7653.23$^{+1083.5}_{-838.5}$ pc \citep{BailerJones2021AJ}. Together with the proper motions from Gaia DR3, we make use of the newly derived distance and heliocentric radial velocity to calculate the Galactic velocity component (U,V,W) and orbital parameters for J1124+4535. The MWPotential2014 from galpy \citep{Bovy2015ApJS} is adopted for the gravitational potential of the MW in order to derive the properties of the orbit, including pericentric radius (R$_{\rm peri}$), apocentric radius (R$_{\rm apo}$), maximum height above the Galactic plane (Z$_{\rm max}$) and eccentricity of the orbit (see Table 1). The results indicate J1124+4535 has a very radial orbit with a eccentricity of 0.53$^{+0.02}_{-0.09}$. The orbit allow J1124+4535 to travel from a 4.2$^{+1.42}_{-0.48}$ kpc area near the Galactic center to a region 13.5$^{+0.94}_{-0.73}$ kpc away from the Galactic center. The bottom panel of Figure 3 shows a Toomre diagram for J1124+4535 and Galactic halo and thick-disk stars from LAMOST DR3 \citep{Xing2018MNRAS}. Interestingly, \citet{Hattori2023ApJ} has analyzed the orbital actions of 161 r-II stars and found that J1124+4535 belongs to a r-II cluster named H22:DTC-17. This cluster is dynamically associated with the LMS-1/Wukong merger group \citep{Malhan2022ApJ} and includes an r-II star associated with the Indus stream, which is suggested to be the remnant of a disrupted dwarf galaxy with mass similar to that of UMi \citep{Ji2020AJ}. The Indus stream, along with the dwarf-galaxy stellar stream LMS-1/Wukong \citep{Yuan2020ApJ,Naidu2020ApJ}, exhibits a clumpy distribution in the orbital action and energy space, suggesting the Indus stream likely originated from the satellite dwarf galaxy of the progenitor LMS-1/Wukong galaxy \citep{Malhan2022ApJ}. \citet{Limberg2023arXiv} determined chemical abundances for 14 stars in LMS-1/Wukong and provided further evidence for this hypothesis. The trends of [Mg/Fe] and [Eu/Fe] versus [Fe/H] in LMS-1/Wukong are shown in Figure 3 for comparison. Figure 4 shows the orbital energy and cylindrical actions of $\sim1700$ r-process-enhanced ([Eu/Fe] $> +0.3$) stars \citep{Shank2023ApJ} and four known MW substructures such as Gaia-Sausage-Enceladus (GSE, \citealt{Helmi2018Natur}), Helmi stream \citep{Helmi1999MNRAS}, LMS-1/Wukong and Thamnos \citep{Koppelman2019AA}. The cluster associated with J1124+4535 is very close to the members of LMS-1/Wukong merger group. The eccentricity (see Figure 4) of this cluster (e = 0.493 $\pm$ 0.097) is also consistent with that of LMS-1/Wukong (e = 0.497 $\pm$ 0.094). The members of this cluster are highly likely to have been brought into the MW by the same merger event involving LMS-1/Wukong merger group. Identifying more stars with a common origin with this cluster will be essential for exploring the properties of their progenitor dwarf galaxy.

\begin{figure*}[h!]
\epsscale{0.9}
\plotone{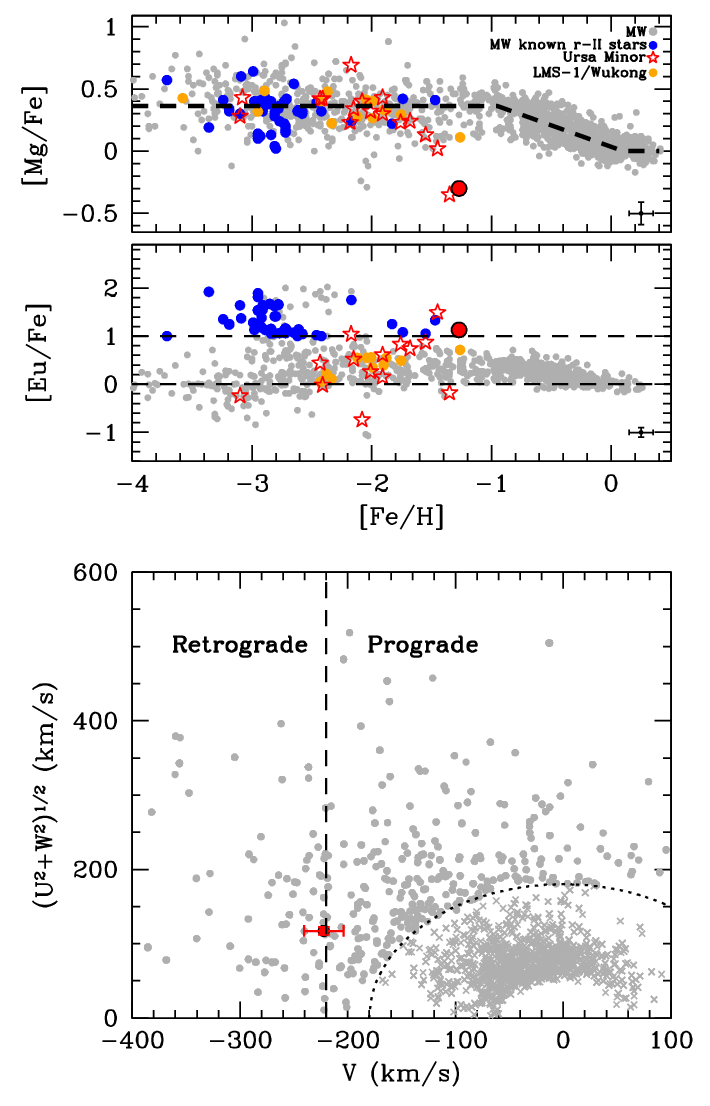}
\caption{Top and middle panels show the [Mg/Fe] and [Eu/Fe] ratios of J1124+4535 (red filled circle) and stars in the classical dwarf galaxy UMi (red open stars) and the stellar stream LMS-1/Wukong (yellow filled circles), in comparison with those of Galactic stars (gray and blue filled circles). The bottom panel is the Toomre diagram for J1124+4535 (red filled circle). The Galactic halo (gray filled circles) and thick-disk stars (black crosses) from LAMOST DR3 \citep{Xing2018MNRAS} are shown for comparison.}
\end{figure*}

\begin{figure*}[h!]
\epsscale{0.9}
\plotone{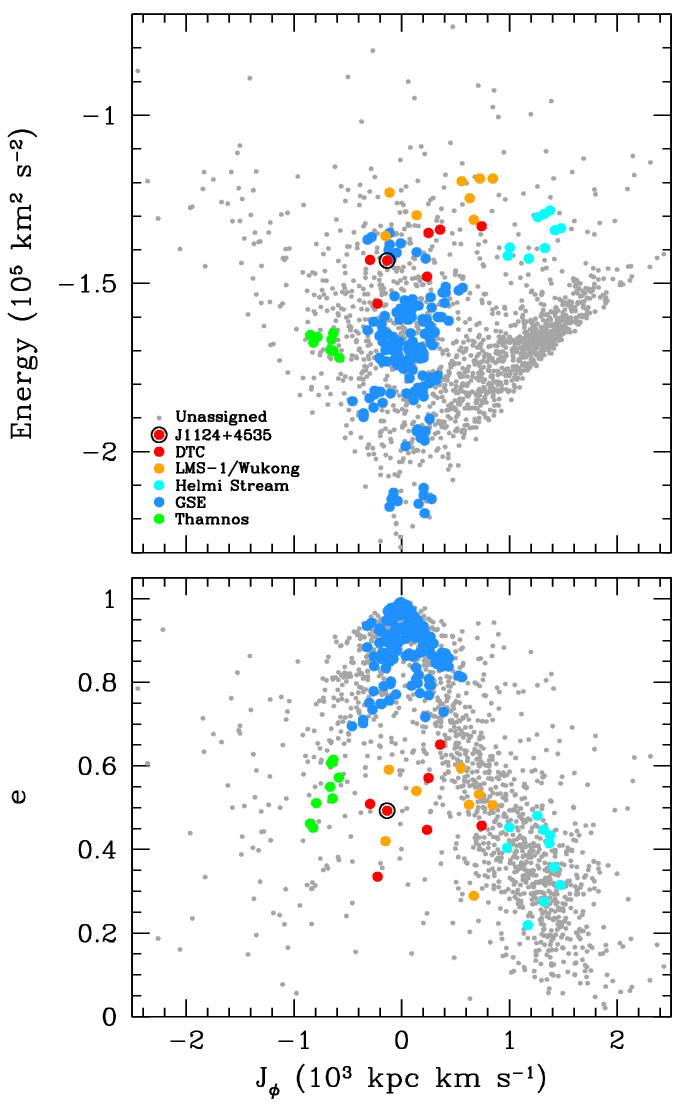}
\caption{Comparing the action, energy, and eccentricities \citep{Shank2023ApJ} of dynamically tagged cluster (DTC) associated with J1124+4535 with the known MW substructures such as Gaia-Sausage-Enceladus (GSE, \citealt{Helmi2018Natur}), Helmi stream \citep{Helmi1999MNRAS}, LMS-1/Wukong and Thamnos \citep{Koppelman2019AA}.}
\end{figure*}


\section{Summary}

We present a detailed abundance analysis of the first $\alpha$-poor r-II star discovered from the LAMOST survey, J1124+4535, based on the newly obtained high-resolution spectrum with Subaru/HDS. The abundances of 12 neutron-capture elements are determined for the analysis of r-process pattern of J1124+4535, including six elements in the second r-process peak and actinide Th. This is the first time we have derived radioactive-decay ages from Th/r abundance ratios for an $\alpha$-poor halo star. The abundance ratios of Th/Eu, Th/Hf, Th/Ir and Th/Os are adopted for the age determination, resulting an age of 11.3 $\pm$ 4.4 Gyr for J1124+4535.

The abundances of light neutron-capture elements (Sr, Y, Zr, Mo, Ru and Rh) in J1124+4535 are relatively low in comparison with the solar r-process pattern while the abundance pattern of heavy elements from Ba to Th closely match the scaled solar r-process pattern. Since the light neutron-capture elements can be produced by the weak r-process, the deviations from the scaled r-process pattern for light neutron-capture elements indicate that the r-process elements in J1124+4535 is dominated by a single r-process event (such as a NSM) corresponding to the main r-process site. In that case, we may conclude J1124+4535 is originated from a dwarf galaxy with a stellar mass $\sim$ 10$^5$M$_\odot$, comparable to the classical dwarf galaxy UMi. Such satellite classical dwarf galaxies are almost as old as the MW. The $\alpha$-poor stars with sub-solar [Mg/Fe] ratios, such as J1124+4535 ([Mg/Fe] = -0.31, [Fe/H] = -1.27), are expected to be formed at the late phase of star formation history of its progenitor dwarf galaxy. The age of J1124+4535 is consistent with this scenario that the star formation in the progenitor dwarf galaxy lasted more than 2 Gyr. More spectroscopic and kinematic analysis of r-process-enhanced stars will help us to identify more accreted components in the Galactic halo.

\acknowledgments

The authors thank the referee for helpful comments and suggestions. The authors also thank Derek Shank for providing the dynamical parameters of r-process-enhanced stars. This work was supported by the National Natural Science Foundation of China grant Nos. 11988101, the JSPS-CAS Joint Research Program, and the science research grants from the China Manned Space Project with No. CMSCSST-2021-B05. Q.X. and H.L. acknowledge support from the Youth Innovation Promotion Association of the CAS (id. 2020058 and Y202017) and the Strategic Priority Research Program of Chinese Academy of Sciences, grant No. XDB34020205. Q.X. is supported by the Beijing Municipal Natural Science Foundation grant No. 1242031. This work was also supported by JSPS KAKENHI Grant Numbers 21H04499, 22K03688, and 23HP8014. This research is based on data collected at Subaru Telescope, which is operated by the National Astronomical Observatory of Japan. We are honored and grateful for the opportunity of observing the Universe from Maunakea, which has the cultural, historical and natural significance in Hawaii.
Guoshoujing Telescope (the Large Sky Area Multi-Object Fiber Spectroscopic Telescope, LAMOST)
is a National Major Scientific Project built by the Chinese Academy of Sciences.
Funding for the project has been provided by the National Development and Reform Commission.
It is operated and managed by the National Astronomical Observatories, Chinese Academy of Sciences.
This work has made use of data from the European Space Agency (ESA) mission
{\it Gaia} (\url{https://www.cosmos.esa.int/gaia}), processed by the {\it Gaia}
Data Processing and Analysis Consortium (DPAC,
\url{https://www.cosmos.esa.int/web/gaia/dpac/consortium}). Funding for the DPAC
has been provided by national institutions, in particular the institutions
participating in the {\it Gaia} Multilateral Agreement.

%

\vspace{5mm}
\facilities{LAMOST, Subaru}


\software{IRAF \citep{Tody1986SPIE,Tody1993ASPC},
          MOOG (v2017; \citealt{Sneden1973PhDT}),
          galpy \citep{Bovy2015ApJS}}



\clearpage

\bibliographystyle{aasjournal}
\bibliography{J1124}



\end{document}